# Contraintes pour modèle et langage multidimensionnels

F. Ghozzi — F. Ravat — O. Teste — G. Zurfluh

*IRIT/SIG*
*Université Paul Sabatier*
*118, Route de Narbonne*
*31062 TOULOUSE cedex 04*
*tel : 05.61.55.74.35*
*fax : 05.61.55.62.58*
*mél : {ghozzi, ravat, teste, zurfluh}irit.fr*

RÉSUMÉ. *Cet article définit un modèle à contraintes pour les bases multidimensionnelles. Notre modèle représente les données en une constellation de faits (sujets d'analyse) associés à des dimensions (axes d'analyse) pouvant être partagées. Chaque dimension peut être structurée en plusieurs hiérarchies (perspectives d'analyse) intégrant différents niveaux de granularités de données. Afin d'assurer la cohérence des données et une plus grande fiabilité de leur manipulation, nous intégrons 5 contraintes sémantiques (exclusion, inclusion, partition, simultanéité, totalité) pouvant être intra-dimension ou inter-dimensions ; les contraintes intra-dimension permettent d'exprimer des contraintes entre les hiérarchies d'une même dimension tandis que les contraintes inter-dimensions portent sur des hiérarchies de dimensions distinctes. Nous étudions également les répercussions de ces contraintes sur les manipulations multidimensionnelles et nous proposons d'étendre les opérateurs multidimensionnels.*

ABSTRACT. *This paper defines a constraint-based model dedicated to multidimensional databases. The model we define represents data through a constellation of facts (subjects of analyse) associated to dimensions (axis of analyse), which are possibly shared. Each dimension is organised according to several hierarchies (views of analyse) integrating several levels of data granularity. In order to insure data consistency, we introduce 5 semantic constraints (exclusion, inclusion, partition, simultaneity, totality) which can be intra-dimension or inter-dimensions; the intra-dimension constraints allow the expression of constraints between hierarchies within a same dimension whereas the inter-dimensions constraints focus on hierarchies of distinct dimensions. We also study repercussions of these constraints on multidimensional manipulations and we provide extensions of the multidimensional operators.*

MOTS-CLÉS : *Bases de données multidimensionnelles, contraintes, opérateurs algébriques.*

# 1. Introduction

Les systèmes d'analyse de données OLAP (*On-Line Analytic Processing*) améliorent les processus décisionnels des organisations (Codd, 1993). De nos jours, ces systèmes OLAP sont le plus souvent composés de bases de données multidimensionnelles permettant une exploitation rapide, efficace et performante des données à des fins décisionnelles (Codd, 1993) (Kimball, 1996).

Cependant, la constitution d'une base de données décisionnelles fiables et cohérentes implique l'intégration de contraintes dans un modèle multidimensionnel (Hurtado *et al*., 2002). Ces contraintes permettent une représentation plus précise de la réalité et interdisent des corrélations incohérentes, offrant ainsi un cadre plus fiable pour les analyses et les prises de décision.

## *1.1. Positionnement de la problématique*

La communauté scientifique a étudié et propose la définition de schémas multidimensionnels organisés en faits à analyser selon différents axes d'analyses appelés dimensions (Kimball, 1996). Les faits sont composés de mesures d'activité et les dimensions comportent des paramètres. Ces derniers sont organisés en hiérarchies représentant différents niveaux de granularité. La manipulation des données multidimensionnelles s'effectuent au travers de cubes de données (Gray *et al.*, 1996), de cubes à *n* dimensions (Gyssen *et al.*, 1997) ou de cubes intégrant une représentation explicite des hiérarchies (Agrawal *et al.*, 1995) (Li *et al.*, 1996) (Lehner, 1998). D'autres modèles gèrent des objets à structures complexes (OOLAP) (Cabibbo *et al.*, 1998) (Trujillo *et al.*, 1998) (Pedersen *et al.*, 1999) et intègrent des fonctionnalités avancées de gestion du temps (Mendelzon *et al.*, 2000).

Cependant, tous ces travaux ne tiennent pas compte de certaines incohérences pouvant survenir lors de l'élaboration et de l'utilisation d'une base de données multidimensionnelles. Par exemple sur une même dimension géographique, la définition d'une hiérarchie organisant les villes en départements et en régions est cohérente pour des villes françaises tandis que cette hiérarchie propre à la France ne peut être appliquée pour des villes situées dans des pays étrangers.

A notre connaissance, seul (Hurtado *et al*., 2002) propose l'intégration de contraintes pour gérer ce type d'incohérences entre les hiérarchies d'une même dimension. Cependant, le langage proposé se focalise sur l'expression de contraintes, et ne permet pas de formaliser les différents types de contraintes. En outre, d'autres conflits peuvent être mis en évidence, notamment entre les hiérarchies de dimensions distinctes. Par exemple, il est impossible d'analyser les ventes de produits classés selon une taxonomie française dans des magasins classés suivant la géographie américaine (états, villes). Ce type d'incompatibilité n'a fait l'objet d'aucune proposition.

*1.2. Contributions et organisation de l'article*

Cet article définit un modèle de données en **constellation** intégrant des **contraintes**. Au niveau des concepts, notre modèle en constellation présente l'avantage de généraliser les approches existantes en autorisant la définition de plusieurs faits (sujets d'analyse) en fonction de dimensions spécifiques et/ou partagées ; en outre, notre approche supporte la multi-hiérarchisation des dimensions offrant ainsi un cadre permettant la définition de plusieurs perspectives d'analyse sur un même axe. De plus, ce modèle conceptuel présente l'avantage de généraliser les approches existantes, tout en étant indépendant des choix d'implantation (ROLAP, MOLAP,…)

Au niveau des contraintes, l'originalité de notre approche réside dans la proposition d'une typologie permettant d'identifier clairement les différentes incohérences pouvant survenir. Cette typologie fait référence à des concepts d'une approche existante (généralisation), mais l'objectif de cet article est de redéfinir des contraintes spécifiques au contexte multidimensionnel (hiérarchies des dimensions). Nos travaux se démarquent de (Hurtado *et al.*, 2002) puisque non seulement nous permettons l'expression de contraintes sur les hiérarchies d'une dimension, mais nous permettons également de définir des contraintes entre différentes dimensions.

L'intégration de contraintes sémantiques dans le modèle multidimensionnel se répercute lors des manipulations multidimensionnelles. Contrairement à la littérature, cette répercussion est étudiée dans cet article ; nous proposons d'étendre les principaux opérateurs utilisés lors des manipulations multidimensionnelles.

La section 2 présente notre modèle multidimensionnel et la définition des contraintes sémantiques. La section 3 étudie les répercutions de ces contraintes sur les opérations multidimensionnelles et propose l'extension des opérateurs.

**2. Modèle multidimensionnel contraint**

Cette section définit notre modèle multidimensionnel basé sur les concepts de constellation (section 2.1), de fait (section 2.2), de dimension multi-instanciable (section 2.3), de hiérarchies multiples (section 2.4) et des contraintes sémantiques d'inclusion, d'exclusion, de simultanéité, de totalité et de partition (section 2.5).

*2.1. Constellation*

Le schéma en constellation est une généralisation du schéma en étoile (Kimball, 1996). Une constellation regroupe plusieurs sujets d'analyse (faits) étudiés selon différents axes (dimensions) éventuellement partagés.

> **Définition.** Une constellation $C$ est définie par ($N^C$, $F^C$, $D^C$, $Star^C$, $Cons^C$) où
>
> – $N^C$ est le nom de la constellation,
>
> – $F^C = \{F_1, F_2,\ldots, F_p\}$ est un ensemble de faits,
>
> – $D^C = \{D_1, D_2,\ldots, D_q\}$ est un ensemble de dimensions,
>
> – $Star^C : F^C \rightarrow 2^{D^C}$ est une fonction associant les faits aux dimensions afin de spécifier les sujets d'analyses et les axes d'étude associés,
>
> – $Cons^C$ représente l'ensemble des contraintes associées à la constellation.

**Exemple.** Un organisme regroupant différentes enseignes d'agences de voyage décide de mettre en place une base de données multidimensionnelles afin d'effectuer l'analyse quotidienne des *ventes* de voyages en fonction du catalogue proposé aux clients dans les différentes agences. La direction souhaite corréler ces analyses avec les *performances* (chiffre d'affaire et marge) des employés de chaque agence. Ces analyses doivent pouvoir être effectuées de manière homogène quelque soit le lieu de vente (France ou Etats-Unis) et le voyage.

Ce besoin peut être traduit par une base de données multidimensionnelles organisée selon une constellation comportant deux faits (*VENTE*, *PERF*) et cinq dimensions (*TEMPS*, *CLIENTS*, *AGENCES*, *EMPLOYES*, *VOYAGES*). La constellation est définie par ($N^C$, $F^C$, $D^C$, $Star^C$, $Cons^C$) où

– $N^C$ = "LOUEVOYAGE",

– $F^C$ = {*VENTES, PERF*},

– $D^C$ = {*TEMPS, EMPLOYES, VOYAGES, AGENCES, CLIENTS*},

– $Star^C$ = { *VENTES*→{ *TEMPS, VOYAGES, AGENCES, CLIENTS* },
        *PERF*→{ *TEMPS, EMPLOYES , AGENCES*} },

– $Cons^C = \{C_1, \ldots \}$. Ces contraintes sont étudiées en section 2.5.

Pour faciliter la compréhension de l'exemple, nous représentons graphiquement la constellation en adaptant le formalisme défini par (Golfarelli *et al.*, 1998).

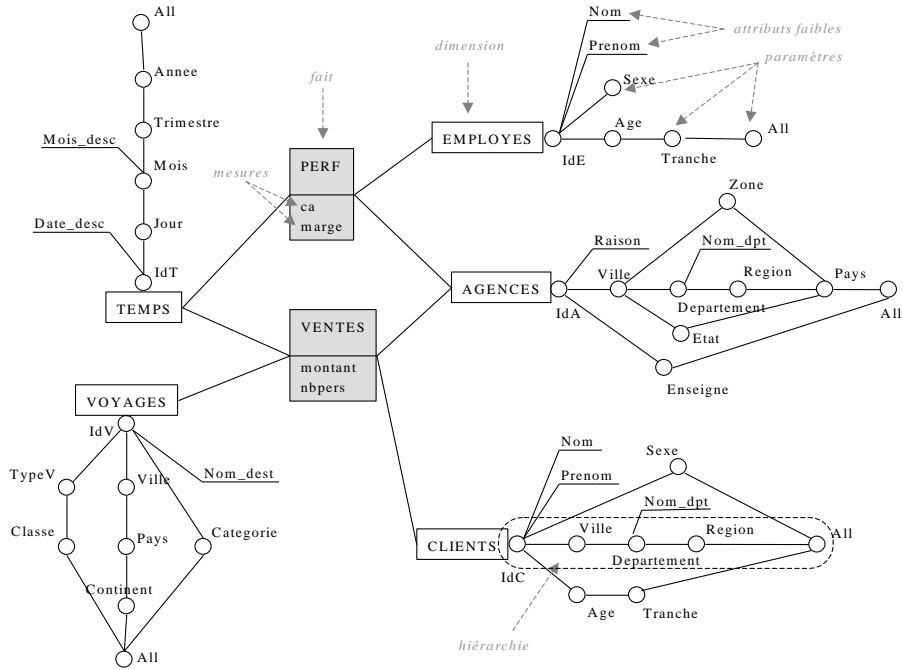

**Figure 1.** Représentation graphique d'une constellation.

## 2.2. Fait

Tout sujet d'analyse est représenté par un fait. Chaque fait est caractérisé par une ou plusieurs mesures représentant les indicateurs analysés.

On pose $DOM = \cup Dom_i$ où chaque $Dom_i$ est un domaine (tels que entier, réel, caractère, chaîne de caractères…), tout élément de $DOM$ est une valeur.

> **Définition.** Un fait $F$ est défini par ($N^F$, $M^F$, $I^F$, $IStar^F$) où
>
> – $N^F$ est le nom du fait,
>
> – $M^F = \{a_1, a_2,…, a_w\}$ est un ensemble de mesures (ou indicateurs),
>
> – $I^F = \{I^F_1, I^F_2,…\}$ est l'ensemble des instances de $F$. Une instance est définie par le n-uplet $[a_1:v_1, a_2:v_2,…, a_w:v_w]$ où $\forall k \in [1..w]$, $a_k \in M^F \wedge v_k \in DOM$.
>
> – $IStar^F$ est une fonction associant chaque instance de $I^F$ à une instance de chaque dimension liée au fait.

**Exemple.** Le fait *VENTES* est constitué de deux mesures (ou indicateurs) représentant le montant de chaque vente de voyage (*montant*) et le nombre de personnes participant au voyage (*nbpers*). Une définition textuelle du fait peut être spécifiée de la manière suivante :

– $N^{VENTES}$ = "*VENTES*",

– $M^{VENTES}$ = { *montant*, *nbpers* },

– $I^{VENTES}$ = { $I^{VENTES}_1$, $I^{VENTES}_2$,… },

– la fonction $IStar^{VENTES}$ est définie par { $I^{VENTES}_1 \rightarrow \{I^{TEMPS}_1, I^{VOYAGES}_1, I^{AGENCES}_1, I^{CLIENTS}_1\}$, $I^{VENTES}_2 \rightarrow \{I^{TEMPS}_1, I^{VOYAGES}_1, I^{AGENCES}_1, I^{CLIENTS}_2\}$,… }.

Notons que les 2 instances de fait précédentes ($I^{VENTES}_1$, $I^{VENTES}_2$) concernent un même voyage ($I^{VOYAGES}_1$) vendu dans une même agence ($I^{AGENCES}_1$), le même jour ($I^{TEMPS}_1$) pour 2 clients différents ($I^{CLIENTS}_1$ et $I^{CLIENTS}_2$). Chaque instance de fait est un n-uplet de la forme suivante :

– $I^{VENTES}_1$ = [ *montant* : 540.00, *nbpers* : 2 ],

– $I^{VENTES}_2$ = [*montant* : 1080.00, *nbpers* : 4 ].

## 2.3. Dimension

Les dimensions représentent les axes d'analyse. Une dimension est formée d'attributs exprimant les caractéristiques en fonction desquelles sont analysées les mesures d'activité. Les attributs d'une dimension peuvent être organisés en hiérarchies, de la granularité la plus fine à la plus générale.

On pose *IDE* un ensemble d'identifiants.

> **Définition.** Une dimension *D* est définie par ($N^D$, $P^D$, $H^D$, $I^D$) où
>
> – $N^D$ est le nom de la dimension,
>
> – $A^D$ = {$a_1, a_2,…, a_u$} est un ensemble d'attributs,
>
> – $H^D$ = {$h^D_1, h^D_2,…, h^D_v$} est un ensemble de hiérarchies,
>
> – $I^D$ = {$I^D_1, I^D_2,…$} est l'ensemble des instances de *D*. Une instance est définie par le n-uplet [$a_1:v_1, …, a_u:v_u$] tel que $\forall k \in [1..u], a_k \in A^D \wedge (v_k \in DOM \vee v_k \in IDE)$.

Toute dimension possède, en plus des attributs définis par l'utilisateur, les attributs *All* et *Id* tels que *dom*(*All*)={'all'} et *dom*(*Id*)∈*IDE* ; *All* désigne la granularité de plus haut niveau tandis que *Id* représente la granularité la plus fine.

**Exemple.** La dimension *AGENCES* peut être définie par :

– $N^{AGENCES}$ = "*AGENCES*",

– $A^{AGENCES}$ = { *IdA, Raison, Ville, Departement, Nom_dpt, Region, Pays, Etat, Zone, Enseigne, All* },

– $H^{AGENCES}$ = { $h^{AGENCES}_1, h^{AGENCES}_2, h^{AGENCES}_3, h^{AGENCES}_4$ },

– $I^{AGENCES}$ = { $I^{AGENCES}_1$, $I^{AGENCES}_2$, $I^{AGENCES}_3$,… }.

Chaque instance est un n-uplet de la forme suivante :

– $I^{AGENCES}_1$ = [ *IdA* : 1, *Raison* : "Agence Campus31", *Ville* : "Toulouse", *Departement* : 31, *Nom_dpt* : "Hte-Garonne", *Region* : "Midi-Pyrénées", *Pays* : "France", *Etat* : NULL, *Zone* : 'S-FR', *Enseigne* : "Fram", *All* : "all" ],

– $I^{AGENCES}_2$ = [ *IdA* : 2, *Raison* : "Agence du Bouchon", *Ville* : "Lyon", *Departement* : 69, *Nom_dpt* : "Rhône", *Region* : "Rhône-Alpes", *Pays* : "France", *Etat* : NULL, *Zone* : 'E-FR', *Enseigne* : "Fram", *All* : "all" ],

– $I^{AGENCES}_3$ = [*IdA* : 2, *Raison* : "Big Appel Agency", *Ville* : "New York", *Departement* : NULL, *Nom_dpt* : NULL, *Region* : NULL, *Pays* : "Etats-Unis", *Etat* : "New York", *Zone* : 'E-US', *Enseigne* : "Travel Express", *All* : "all"].

*2.4. Hiérarchie*

Les hiérarchies permettent de déterminer les niveaux de granularité auxquels peuvent être manipulées les mesures des faits. Chaque hiérarchie propose un ordonnancement des attributs d'une dimension afin de spécifier ces différents niveaux de granularité.

> **Définition.** Une hiérarchie $h^D_i$ est un chemin élémentaire acyclique débutant par *Id* et se terminant par *All*. Elle est définie par ($N^{hDi}$, $Param^{hDi}$, $Suppl^{hDi}$, $Cond^{hDi}$) où
>
> – $N^{hDi}$ est le nom de la hiérarchie,
>
> – $Param^{hDi}$ : $P \to P$ ($P \subseteq A^D$) est une fonction décrivant la hiérarchie des attributs (chaque attribut est appelé paramètre de la hiérarchie),
>
> – $Suppl^{hDi}$ : $P \to 2^{(AD-P)}$ est une fonction spécifiant les attributs faibles qui complètent la sémantique des paramètres (chaque paramètre est associé à un ensemble d'attributs faibles),
>
> – $Cond^{hDi}$ est une expression booléenne définissant la condition d'appartenance des instances de la dimension à une hiérarchie.

Nous notons $I^D_k \in h^D_i$ pour indiquer qu'une instance $I^D$ satisfait la condition $Cond^h$ et par conséquent $I^D_k$ appartient à la hiérarchie $h^D_i$.

**Exemple.** Pour compléter la définition de la dimension *AGENCES*, nous définissons les 4 hiérarchies suivantes :

– $H^{AGENCES}_1$ = ("geo_fr", {*IdA→Ville*, *Ville→Departement*, *Departement→Region*, *Region→Pays*, *Pays→All*}, {*IdA→{Raison}*, *Departement→{Nom_dpt}*}, *Pays* = "France" ),

– $h^{AGENCES}_2$ = ("geo_us", {*IdA→Ville*, *Ville→Etat*, *Etat→Pays*, *Pays→All*}, {*IdA→{Raison}*}, *Pays* = "Etats-Unis" $\wedge$ *Etat* $\neq$ NULL ),

– $h^{AGENCES}_3$ = ("geo_zn", {*IdA→Ville*, *Ville→Zone*, *Zone→Pays*, *Pays→All*}, {*IdA→{Raison}*}, *Zone* $\neq$ NULL ),

– $h^{AGENCES}_4$ = ("ens", {$IdA{\rightarrow}Enseigne$, $Enseigne{\rightarrow}All$}, {$IdA{\rightarrow}${$Raison$}}, $Enseigne \neq$ NULL ).

La hiérarchie $H^{AGENCES}_1$ décrit les instances suivant l'organisation géographique de la France en ville, département et région tandis que la hiérarchie $h^{AGENCES}_2$ est relative à l'organisation géographique des Etats-Unis. La hiérarchie $h^{AGENCES}_3$, commune à la France et aux Etats-Unis décrit la position des villes selon l'indication nord, sud, est, ouest. La hiérarchie $h^{AGENCES}_4$ décrit chaque agence en fonction de son appartenance à une enseigne spécifique.

La spécificité de multi-instanciation de notre modèle réside dans l'intégration d'une condition d'appartenance des instances de la dimension aux hiérarchies. Ainsi, les instances {$I^{AGENCES}_1$, $I^{AGENCES}_2$} appartiennent à $h^{AGENCES}_1$ tandis que l'instance {$I^{AGENCES}_3$} appartient à $h^{AGENCES}_2$ et les instances {$I^{AGENCES}_1$, $I^{AGENCES}_2$, $I^{AGENCES}_3$} appartiennent à $h^{AGENCES}_3$ et à $h^{AGENCES}_4$.

**Exemple.** On considère la dimension *VOYAGES* représentant les voyages proposés par l'organisme. Nous définissons ses hiérarchies de la manière suivante :

– $h^{VOYAGES}_1$ = {"cla_us", {$IdV{\rightarrow}TypeV$, $TypeV{\rightarrow}Class$, $Class{\rightarrow}All$}, $Class \neq$ NULL $\wedge$ $TypeV \neq$ NULL),

– $h^{VOYAGES}_2$ = {"cla_int", {$IdV{\rightarrow}Ville$, $Ville{\rightarrow}Pays$, $Pays{\rightarrow}Continent$, $Continent{\rightarrow}All$}, TRUE),

– $h^{VOYAGES}_3$ = {"cla_fr", {$IdV{\rightarrow}Categorie$, $Categorie{\rightarrow}All$}, $Categorie \neq$ NULL).

La hiérarchie $h^{VOYAGES}_1$, spécifique aux Etats-Unis décrit les voyages suivant une classification en type de voyage, puis en classe de voyage tandis que la hiérarchie $h^{VOYAGES}_3$ décrit une nomenclature inhérente à la France. La hiérarchie $h^{VOYAGES}_2$ offre une vision commune à toutes les instances décrivant la destination du voyage.

*2.5. Contraintes*

Les travaux exposés dans cet article se focalisent sur les contraintes sémantiques. Ces contraintes agissent sur la cohérence et la manipulation des données entreposées. Nous identifions des besoins non pris en compte dans les modèles actuels.

– Un premier besoin concerne la nécessité d'exprimer des interactions entre les différentes perspectives (hiérarchies) d'un même axe d'analyse (dimension). Ainsi, les instances relatives à une hiérarchie décrivant la géographie française ne peuvent être décrites suivant une hiérarchie relative à la géographie américaine. Ceci peut être traduit par des **contraintes intra-dimension** portant sur les hiérarchies d'une même dimension.

– Un second besoin est relatif à la possibilité de spécifier selon quels axes (dimensions) et/ou quelles perspectives (hiérarchies) peuvent être associées les mesures d'activité (mesures d'un fait). Cet aspect est pris en compte par des contraintes sur les hiérarchies de dimensions distinctes que nous appelons **contraintes inter-dimensions**.

### 2.5.1. Contraintes intra-dimensions

Les contraintes intra-dimensions sont exprimées entre les hiérarchies d'une même dimension. Il s'agit de contraintes portant sur les sous-ensembles d'instances de la dimension définis par la condition d'appartenance associée à chacune des hiérarchies de la dimension. Nous définissons 5 contraintes intra-dimensions : l'exclusion, l'inclusion, la simultanéité, la partition et la totalité.

On pose $D$ une dimension et $h_1 \in H^D$, $h_2 \in H^D$ deux hiérarchies de la dimension.

#### 2.5.1.1. Exclusion intra-dimension

**Définition**. L'exclusion, notée $\otimes$, entre deux hiérarchies d'une dimension traduit qu'une instance de la dimension appartenant à une hiérarchie (l'instance satisfait la condition d'appartenance associée à la hiérarchie) n'appartient pas à la seconde hiérarchie et réciproquement.

$h_1 \otimes h_2$ ssi $\forall I^D_{k1} \in h_1 \land \forall I^D_{k2} \in h_2 \Rightarrow I^D_{k1} \neq I^D_{k2}$.

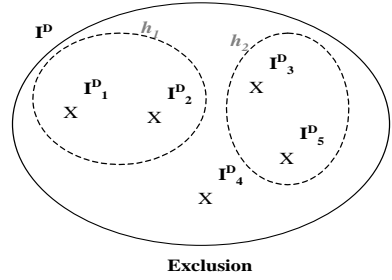

**Exclusion**

#### 2.5.1.2. Inclusion intra-dimension

**Définition**. L'inclusion, notée $\odot$, entre deux hiérarchies d'une dimension traduit que toutes les instances de la dimension appartenant à une première hiérarchie appartiennent à la seconde hiérarchie.

$h_1 \odot h_2$ ssi $\forall I^D_k \in h_1 \Rightarrow I^D_k \in h_2$.

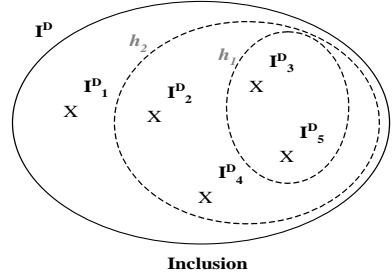

**Inclusion**

#### 2.5.1.3. Simultanéité intra-dimension

**Définition**. La simultanéité, notée $\ominus$, entre deux hiérarchies d'une dimension traduit que toutes les instances de la dimension appartenant à une première hiérarchie appartiennent alors à la seconde et réciproquement (double inclusion).

$h_1 \ominus h_2$ ssi $\forall I^D_k \in h_1 \Leftrightarrow I^D_k \in h_2$.

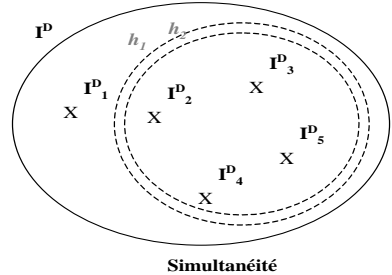

**Simultanéité**

2.5.1.4. Totalité intra-dimension

**Définition**. La totalité, notée ⊖, entre deux hiérarchies d'une dimension traduit que toute instance de la dimension appartient à l'une ou (non exclusif) l'autre des hiérarchies. Ainsi, toute instance de la dimension appartient à l'une des deux hiérarchies et éventuellement aux deux hiérarchies.

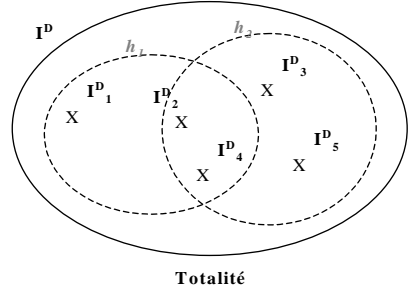

Totalité

$h_1 \ominus h_2$ ssi $\forall I^D_k \in I^D, I^D_k \in h_1 \vee I^D_k \in h_2$.

2.5.1.5. Partition intra-dimension

**Définition**. La partition, notée ⊘, entre deux hiérarchies d'une dimension traduit que toute instance de la dimension appartient à l'une ou (exclusif) l'autre des hiérarchies. Ainsi, toute instance de la dimension appartient obligatoirement à l'une des deux hiérarchies, mais pas aux deux.

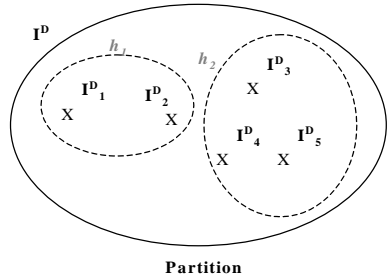

Partition

$h_1 \oslash h_2$ ssi $(\forall I^D_k \in I^D, I^D_k \in h_1 \vee I^D_k \in h_2) \wedge (\forall I^D_{k1} \in h_1 \wedge \forall I^D_{k2} \in h_2 \Rightarrow I^D_{k1} \neq I^D_{k2})$.

**Propriété**. On peut noter que la contrainte de partition intra-dimension revient à exprimer entre les deux hiérarchies les contraintes de totalité et d'exclusion intra-dimension.

2.5.1.6. Exemples d'utilisation des contraintes intra-dimension

Dans l'exemple présenté en figure 1, les agences de voyage sont situées soit aux Etats-Unis, soit en France. En outre, les agences situées aux Etats-Unis ne peuvent pas être analysées selon la hiérarchie relative à la géographie française, et réciproquement. De plus, toute agence est localisée suivant sa zone géographique dans le pays (Nord, Sud, Est, Ouest) et appartient à une enseigne.

Pour compléter la sémantique de la constellation, nous modélisons les contraintes de partition, de simultanéité et d'inclusion intra-dimension (sur *AGENCES*) suivantes :

– $h^{AGENCES}_1 \oslash h^{AGENCES}_2$ (partition),
– $h^{AGENCES}_3 \ominus h^{AGENCES}_4$ (simultanéité),
– $h^{AGENCES}_1 \odot h^{AGENCES}_3$, $h^{AGENCES}_2 \odot h^{AGENCES}_3$, $h^{AGENCES}_1 \odot h^{AGENCES}_4$ et $h^{AGENCES}_2 \odot h^{AGENCES}_4$ (inclusions).

*2.5.2. Contraintes inter-dimensions*

Les contraintes inter-dimensions sont exprimées entre les hiérarchies de dimensions distinctes associées à un même fait. Il s'agit donc de contraintes relatives aux instances du fait associées aux instances de dimension.

On pose $D_1$ et $D_2$ deux dimensions liées à un fait $F$ ($D_1 \in Star^C(F) \wedge D_2 \in Star^C(F)$) et $h_1 \in H^{D1}$, $h_2 \in H^{D2}$ deux hiérarchies définies sur les dimensions $D_1$ et $D_2$.

2.5.2.1. Exclusion inter-dimensions

**Définition**. L'exclusion entre deux hiérarchies de dimensions distinctes traduit qu'une instance de fait liée à une instance de la dimension appartenant à une hiérarchie n'est pas liée à une instance de dimension appartenant à la seconde hiérarchie et réciproquement.

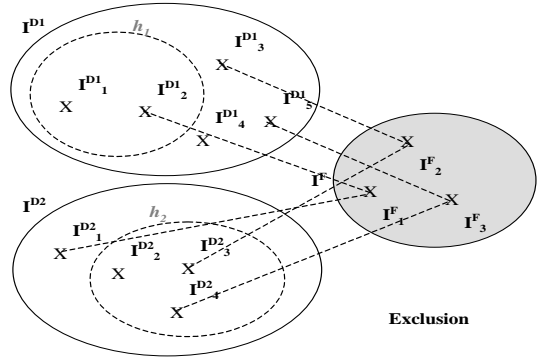

Exclusion

$h_1 \otimes h_2$ ssi $(\forall I^F_j \in I^F \mid \exists I^{D1}_{k1} \in h_1 \wedge I^{D1}_{k1} \in IStar^F(I^F_j)) \Rightarrow \neg(\exists I^{D2}_{k2} \in h_2 \mid I^{D2}_{k2} \in IStar^F(I^F_j))$.

2.5.2.2. Inclusion inter-dimensions

**Définition**. L'inclusion entre deux hiérarchies de dimensions distinctes traduit que toutes les instances du fait liées aux instances de la dimension appartenant à une première hiérarchie sont également liées aux instances appartenant à la seconde hiérarchie.

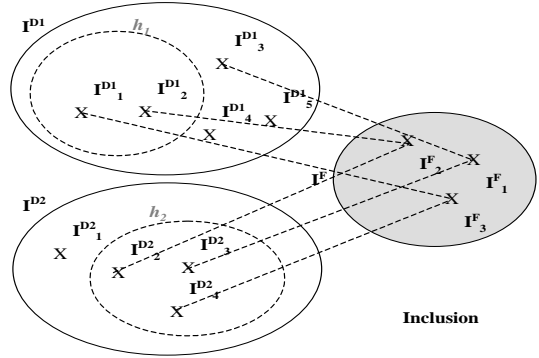

Inclusion

$h_1 \odot h_2$ ssi $(\forall I^F_j \in I^F \mid \exists I^{D1}_{k1} \in h_1 \wedge I^{D1}_{k1} \in IStar^F(I^F_j)) \Rightarrow \exists I^{D2}_{k2} \in h_2 \mid I^{D2}_{k2} \in IStar^F(I^F_j)$

2.5.2.3. Simultanéité inter-dimensions

**Définition**. La simultanéité entre deux hiérarchies de dimensions distinctes traduit que toutes les instances du fait liées aux instances de la dimension appartenant à une première hiérarchie sont également liées aux instances appartenant à la seconde et réciproquement (double

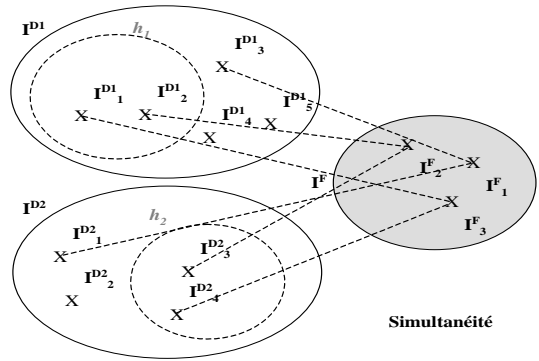

Simultanéité

inclusion).

$$h_1 \ominus h_2 \text{ ssi } (\forall I^F_j \in I^F \mid \exists I^{D1}_{k1} \in h_1 \wedge I^{D1}_{k1} \in IStar^F(I^F_j)) \Leftrightarrow \exists I^{D2}_{k2} \in h_2 \mid I^{D2}_{k2} \in IStar^F(I^F_j)$$

2.5.2.4. Totalité inter-dimensions

**Définition**. La totalité entre deux hiérarchies de dimensions distinctes traduit que toute instance du fait est liée à une instance appartenant à l'une des deux hiérarchies et éventuellement aux deux hiérarchies.

$h_1 \ominus h_2$ ssi $\forall I^F_j \in I^F$, $(\exists I^{D1}_{k1} \in h_1 \mid I^{D1}_{k1} \in IStar^F(I^F_j)) \vee (\exists I^{D2}_{k2} \in h_2 \mid I^{D2}_{k2} \in IStar^F(I^F_j))$

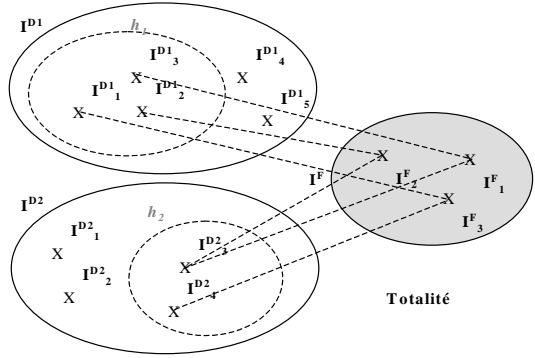

Totalité

2.5.2.5. Partition inter-dimensions

**Définition**. La partition entre deux hiérarchies de dimensions distinctes traduit que toute instance du fait est liée à une instance de dimension appartenant à l'une des deux hiérarchies, mais pas aux deux, ni à aucune des deux.

$h_1 \oslash h_2$ ssi $(\forall I^F_j \in I^F, (\exists I^{D1}_{k1} \in h_1 \mid I^{D1}_{k1} \in IStar^F(I^F_j)) \vee (\exists I^{D2}_{k2} \in h_2 \mid I^{D2}_{k2} \in IStar^F(I^F_j))) \wedge ((\forall I^F_j \in I^F \mid \exists I^{D1}_{k1} \in h_1 \wedge I^{D1}_{k1} \in IStar^F(I^F_j)) \Rightarrow \neg(\exists I^{D2}_{k2} \in h_2 \mid I^{D2}_{k2} \in IStar^F(I^F_j)))$.

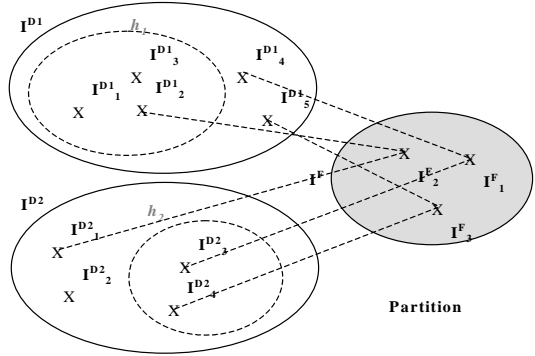

Partition

**Propriété**. On peut noter que la contrainte de partition inter-dimensions revient à exprimer entre les deux hiérarchies les contraintes de totalité inter-dimensions et d'exclusion inter-dimensions.

2.5.2.6. Exemples d'utilisation des contraintes inter-dimensions

Dans l'exemple présenté en figure 1, tout voyage vendu par une agence américaine est classé suivant une nomenclature spécifique aux Etats-Unis ($h^{VOYAGES}_1$) tandis que tout voyage vendu en France est organisé selon la classification définie par la hiérarchie $h^{VOYAGES}_3$. De plus, tous les voyages peuvent être analysés en fonction de leur destination suivant la hiérarchie $h^{VOYAGES}_2$.

Pour compléter la sémantique de la constellation, nous modélisons les contraintes de partition, de simultanéité et d'inclusion suivantes :

- $h^{AGENCES}_1 \oslash h^{VOYAGES}_3$ et $h^{AGENCES}_2 \oslash h^{VOYAGES}_1$ (partitions),
- $h^{AGENCES}_1 \ominus h^{VOYAGES}_1$ et $h^{AGENCES}_2 \ominus h^{VOYAGES}_3$ (simultanéités),
- $h^{AGENCES}_1 \odot h^{VOYAGES}_2$, $h^{AGENCES}_2 \odot h^{VOYAGES}_2$, … (inclusions).

**3. Extensions des opérateurs algébriques OLAP**

Les contraintes sémantiques intra-dimension et inter-dimensions agissent sur les opérations multidimensionnelles (Ravat *et al.*, 2001). Dans les sections suivantes, nous étudions l'impact de ces contraintes sur les principaux opérateurs OLAP. La prise en compte des contraintes nécessite l'extension des opérateurs multidimensionnels afin de permettre ou non le maintien de l'analyse lors de manipulations mettant en jeu des données pour lesquelles des contraintes sémantiques sont définies.

Nous utilisons la notation [ ] pour indiquer un paramètre optionnel dans la définition des opérateurs.

*3.1. Préliminaires*

La visualisation d'une constellation est centrée sur un fait au travers d'une table dimensionnelle. Le choix de visualiser en détail les données d'un seul fait réduit la complexité de l'information visualisée et donc facilite l'interprétation et l'analyse des données. La représentation sous forme de table est motivée par sa simplicité et sa précision ; il s'agit d'une représentation très répandue à laquelle les décideurs sont habitués (Agrawal *et al.*, 1995) (Gyssen *et al.*, 1997) notamment au travers des différents outils du marché.

**Définition**. Une table dimensionnelle $TD^{Fk}$ est définie par ($F_k$, <$D_1, D_2, …, D_v$>, $h^{D1}_i$, $h^{D2}_j$, $param^{hi}$, $param^{hj}$, $Pred^D$) où

- $F_k$ est le fait visualisé,
- <$D_1, D_2, …, D_v$>$\subseteq Star^C(F_k)$ est la liste des dimensions associées au fait $F_k$ avec $D_1$ et $D_2$ les dimensions courantes utilisées pour visualiser les valeurs des mesures du fait ($D_1$ en ligne, $D_2$ en colonne),
- $h^{D1}_i \in H^{D1}$ et $h^{D2}_j \in H^{D2}$ sont les deux hiérarchies utilisées pour visualiser les données des dimensions courantes,
- $param^{hi}$ et $param^{hj}$ désignent la liste des paramètres affichés de chaque dimension,
- $Pred^D$ est un ensemble de prédicats portant sur les dimensions.

**Définition**. L'opérateur de visualisation d'une constellation est défini par l'expression Display($C$, $F_k$, $D_m$, $D_n$, $h^{Dm}_i$, $h^{Dn}_j$) = $TD^{Fk}$ où

- $C$ désignent une constellation,

– $F_k \in F^C$ est le fait de la constellation visualisé,

– $D_m \in Star^C(F_k)$ et $D_n \in Star^C(F_k)$ sont deux dimensions liées au fait $F_k$,

– $h^{Dm}_i \in H^{Dm}$ et $h^{Dn}_j \in H^{Dn}$ sont deux hiérarchies des dimensions $D_m$ et $D_n$,

– $TD^{Fk} = (F_k, <D_m, D_n,...>, h^{Dm}_i, h^{Dn}_j, <p_s>, <p_t>, \emptyset)$ est la table dimensionnelle résultat | $Param^{hDmi}(p_s)=All$ et $Param^{hDnj}(p_t)=All$.

**Exemple**. On considère le schéma présenté à la figure 1. Un utilisateur souhaite visualiser les ventes réalisées chaque année dans les agences françaises. Il définit l'expression suivante pour construire la table dimensionnelle correspondant au besoin d'analyse.

Display(*VENTES*, *AGENCES*, *TEMPS*, *geo_fr*, *h_an*) = $T^{VENTES}$

Par abus de notation, on désigne les éléments de la base (constellation, faits, dimensions…) par leur nom. Le résultat de l'opération est visualisé sous la forme d'une table dimensionnelle $T^{VENTES}$ = (*VENTES*, <*AGENCES*, *TEMPS*, *VOYAGES*, *CLIENTS*>, *geo_fr*, *h_an*, <*pays*>, <*annee*>, $\emptyset$) représentée dans la figure 2. Cette table permet d'étudier les montants et le nombre moyen de personnes effectuants des voyages. L'analyse est effectuée en fonction des années et des pays de vente (quelque soit le client et le voyage).

| VENTES | | AGENCES.geo_fr | |
|---|---|---|---|
| (montant, nbpers) | | pays | France |
| TEMPS.h_an | annee | | |
| | 2000 | | (500.00, 4) |
| | 2001 | | (800.00, 4) |
| | 2002 | | (1200.00, 5) |
| CLIENTS.All = 'all' | | | |
| VOYAGES.All = 'all' | | | |

**Figure 2 :** Exemple de table dimensionnelle.

La représentation des constellations sous forme de tables dimensionnelles offre une visualisation simple et précise aux analystes et aux décideurs. Ces tables offrent un cadre adapté pour l'application d'opérations dites multidimensionnelles. Parmi les opérations les plus importantes, nous pouvons citer les forages (section 3.2) et les rotations (section 3.3).

### 3.2. Opérations de forages (RollUp / DrillDown)

Les opérations de forage sont parmi les plus importantes du processus OLAP. Elles permettent d'augmenter ou de diminuer ("*forer*") le détail avec lequel l'information est représentée. Ces opérations de forage manipulent les différents niveaux de granularité des données multidimensionnelles ; plus précisément, le forage consiste à ajouter ou supprimer les paramètres d'une hiérarchie associée à une dimension lors de la visualisation des données.

Les contraintes que nous avons définies s'expriment entre hiérarchies tandis que l'opération de forage consiste à naviguer le long d'une hiérarchie particulière. Ainsi, il n'est pas utile d'étendre ces opérateurs.

Toutefois, nous présentons un exemple comparatif, avec et sans contraintes, de résultats d'opérateurs de manipulation multidimensionnelle.

**Exemple.** Nous considérons la constellation dont le schéma est décrit par la figure 1. Comme dans l'exemple précédent, un analyste souhaite visualiser les ventes (montant des ventes et nombre de personnes participant au voyage vendu) en fonction des agences et du temps. Afin d'affiner l'analyse, le décideur veut visualiser l'axe des agences au niveau de granularité '*Region*' ; une opération de forage vers le bas doit être appliquée sur la dimension *AGENCES*.

Le résultat souhaité peut être obtenue à l'aide de l'expression suivante : DrillDown(Display(*VENTES*, *AGENCES*, *TEMPS*, *geo_fr*, *h_an*), *AGENCES*, *Region*).

La figure 3 présente le résultat obtenu lorsque l'on considère une constellation sans contrainte. Dans ce cas, l'ensemble des instances de la dimension *AGENCES* sont visualisées en fonction de la hiérarchie choisie ; dans l'exemple, l'analyste visualise les agences suivant la hiérarchie décrivant l'organisation géographique de la France. Cette vision de la dimension *AGENCES* est cohérente pour les instances d'agences situées en France, mais elle s'avère incohérente pour les agences localisées dans d'autres pays tels que les Etats-Unis ; une valeur NULL est alors obtenue.

| VENTES (montant, nbpers) | | AGENCES.geo_fr | | | | |
|---|---|---|---|---|---|---|
| | | pays | France | | | Etats-Unis |
| | | region | Midi-Pyrénées | Gironde | Languedoc-R | NULL |
| TEMPS.h_an | annee | | | | | |
| | 2000 | | (300.00, 5) | (150.00, 4) | (50.00, 2) | (700.00, 5) |
| | 2001 | | (400.00, 4) | (250.00, 4) | (150.00, 3) | (850.00, 5) |
| | 2002 | | (600.00, 5) | (400.00, 5) | (200.00, 2) | (1100.00, 4) |
| CLIENTS.All = 'all' | | | | | | |
| VOYAGES.All = 'all' | | | | | | |

**Figure 3 :** Exemple de table dimensionnelle sans contrainte.

La figure 4 décrit le résultat obtenu dans le cadre d'une constellation modélisée avec des contraintes, conformément au modèle que nous proposons. Dans ce cas, les instances de la dimension *AGENCES* sont associées aux hiérarchies qui donnent une vision cohérente des données ; autrement dit, les agences localisées en France sont appartiennent à la hiérarchie *geo_fr* tandis que les agences localisées aux Etats-Unis n'appartiennent pas à cette hiérarchie. Ainsi, lors de la visualisation des agences suivant la description géographique française, seules les agences situées en France sont prises en compte dans la table dimensionnelle.

| VENTES (montant, nbpers) | | AGENCES.geo_fr | | | |
|---|---|---|---|---|---|
| | | pays | France | | |
| | | region | Midi-Pyrénées | Gironde | Languedoc-R. |
| TEMPS.h_an | annee | | | | |
| | 2000 | | (300.00, 5) | (150.00, 4) | (50.00, 2) |
| | 2001 | | (400.00, 4) | (250.00, 4) | (150.00, 3) |
| | 2002 | | (600.00, 5) | (400.00, 5) | (200.00, 2) |
| CLIENTS.All = 'all' | | | | | |
| VOYAGES.All = 'all' | | | | | |

**Figure 4 :** Exemple de table dimensionnelle avec contraintes.

Si l'analyste souhaite visualiser toutes les agences entreposées dans la constellation, il doit choisir une hiérarchie cohérente avec tous les enregistrements des agences. Il peut, par exemple, se positionner sur la hiérarchie *geo_zn* décrivant la localisation des agences en fonction d'une zone géographique ; voir l'exemple suivant de la section 3.3.1.

### *3.3. Opérations de rotations*

Les opérations de rotation, classiques en analyse multidimensionnelle, permettent d'analyser les données selon différents axes et perspectives. Cette opération s'applique aux dimensions et aux hiérarchies ; elle consiste à permuter deux dimensions ou deux hiérarchies afin de changer les paramètres utilisés pour visualiser les mesures d'activité.

#### *3.3.1. Rotation des hiérarchies (HRotate)*

L'opération de rotation des hiérarchies, notée *HRotate*, consiste à changer la perspective d'analyse sur une même dimension. Dans le cadre de la multi-instanciation, la rotation implique deux possibilités :

– réinitialiser l'analyse en utilisant l'ensemble des instances associées à la nouvelle hiérarchie,

– maintenir l'analyse en utilisant le même ensemble d'instances.

**Définition**. L'opérateur de rotation de hiérarchies se définit par l'expression HRotate($TD^{Fk}$, $D_m$, $h^{Dm}_x$, $h^{Dm}_y$ [, *flag* ]) = $TD'^{Fk}$ où

– $TD^{Fk}$ et $TD'^{Fk}$ désignent les tables dimensionnelles origine et résultat,

– $D_m$ est une dimension liée au fait visualisé $F_k$,

– $h^{Dm}_{x \in \{i, j\}} \in H^{Dm}$ et $h^{Dm}_y \in H^{Dm}$ sont deux hiérarchies de la dimension $D_m$,

– *flag* est un paramètre optionnel indiquant si l'opération est réalisée en maintenant l'analyse (*flag=true*) ou non (*flag=false*, valeur par défaut),

– si *x=i* alors $TD'^{Fk} = (F_k, <D_m, D_n,...>, h^{Dm}_y, h^{Dn}_j, <p_s>, <p_t>, \varnothing)$ sinon $(F_k, <D_m, D_n,...>, h^{Dm}_i, h^{Dn}_y, <p_s>, <p_t>, \varnothing)$.

**Exemple.** On considère le schéma présenté à la figure 1 et la représentation en table dimensionnelle de la figure 2. Un utilisateur peut souhaiter changer la hiérarchie *geo_fr* de la dimension *AGENCES* afin de visualiser les données selon la perspective offerte par la hiérarchie *geo_zn* (représentation des agences suivant la répartition en zones). Cette opération peut être exprimée par l'expression algébrique suivante :

HRotate($T^{VENTES}$, *AGENCES*, *geo_fr*, *geo_zn*, *false*) = $T'^{VENTES}$

Le résultat de l'opération est visualisé sous la forme d'une table dimensionnelle représentée dans la figure 5. Cette rotation s'applique sur des hiérarchies pour lesquelles une contrainte d'inclusion est définie (*geo_fr*⊚*geo_zn*) puisque les instances appartenant à la hiérarchie *geo_fr* sont incluses dans la hiérarchie *geo_zn*. Lors du passage d'une hiérarchie englobée (*geo_fr*) à une hiérarchie englobante (*geo_zn*) nous avons choisi de réinitialiser l'analyse en affichant toutes les instances appartenant à la nouvelle hiérarchie (*flag=false*).

| VENTES | | AGENCES.geo_zn | | |
|---|---|---|---|---|
| (montant, nbpers) | | pays | France | Etats-Unis |
| TEMPS.h_an | annee | | | |
| | 2000 | | (500.00, 4) | (700.00, 5) |
| | 2001 | | (800.00, 4) | (850.00, 5) |
| | 2002 | | (1200.00, 5) | (1100.00, 4) |
| CLIENTS.All = 'all' | | | | |
| VOYAGES.All = 'all' | | | | |

**Figure 5 :** Résultat de l'opération de rotation de hiérarchies.

*3.3.2. Rotation des dimensions (DRotate)*

L'opération de rotation des dimensions, notée *DRotate*, permet de modifier un axe d'analyse, c'est à dire de changer une dimension parmi celles utilisées pour visualiser les données. De manière analogue à la rotation des hiérarchies, cette opération implique soit la réinitialisation de l'analyse (nouvel ensemble d'instances), soit le maintien de l'analyse (même ensemble d'instances).

**Définition**. L'opérateur de rotation de dimensions se définit par l'expression DRotate($TD^{Fk}, D_x, D_y, h^{Dy}$ [, *flag* ]) = $TD'^{Fk}$ où

– $TD^{Fk}$ et $TD'^{Fk}$ désignent les tables dimensionnelles origine et résultat,

– $D_{x \in \{m, n\}}$ est une dimension liée au fait visualisé $F_k$,

– $D_j$ est la nouvelle dimension utilisée pour visualiser les données de $F_k$,

– $h^{Dy} \in H^{Dy}$ est une hiérarchie définie sur la dimension $D_y$,

– *flag* est un paramètre optionnel indiquant si l'opération est réalisée en maintenant l'analyse (*flag=true*) ou non (*flag=false*, valeur par défaut).

– si $x=m$ alors $TD'^{Fk} = (F_k, <D_y, D_n,...>, h^{Dy}, h^{Dn}_j, <p_s>, <p_t>, \varnothing)$ sinon $(F_k, <D_m, D_y,...>, h^{Dm}_i, h^{Dy}, <p_s>, <p_t>, \varnothing)$.

**Exemple.** On considère la représentation en table dimensionnelle de la figure 5. L'utilisateur souhaite poursuivre l'analyse en effectuant une visualisation des ventes non plus par la dimension *AGENCES* mais en utilisant la dimension *VOYAGES*. Cette opération peut être exprimée par l'expression algébrique suivante :

DRotate($T'^{VENTES}$, *AGENCES*, *VOYAGES*, *cla_int*, *true*) = $T''^{VENTES}$

Cette rotation s'applique sur des dimensions dont les hiérarchies sont en contrainte d'inclusion (*geo_fr*⊙*cla_int*) puisque toutes les instances de *VENTES* liées à une instance de dimension appartenant à la hiérarchie *geo_fr* (de la dimension *AGENCES*) sont liées à une instance de la hiérarchie *cla_int* (de la dimension *VOYAGES*). L'utilisateur maintient l'analyse sur les instances concernant les agences françaises en utilisant l'option *flag=true*. Ainsi, les données visualisées dans la table dimensionnelle représentée dans la figure 6 concernent les voyages vendus par les agences françaises.

| VENTES | VOYAGES.cla_int | | | |
|---|---|---|---|---|
| (montant, nbpers) | **continent** | Europe | Amérique | Afrique |
| **TEMPS**.h_an **annee** | | | | |
| 2000 | | (200.00, 4) | (170.00, 4) | (130.00, 4) |
| 2001 | | (500.00, 5) | (200.00, 4) | (100.00, 3) |
| 2002 | | (800.00, 6) | (250.00, 5) | (150.00, 4) |
| CLIENTS.All = 'all' | | | | |
| AGENCES.Pays = 'France' | | | | |

**Figure 6 :** Résultat de l'opération de rotation de dimensions.

Comme toute algèbre, notre proposition permet d'exprimer des requêtes complexes en combinant les différents opérateurs (Ravat *et al*, 2002).

**Exemple.** Un décideur souhaite, à partir de l'exemple précédent, visualiser les ventes des voyages suivant la nomenclature française (classe et type) et par enseigne.

DrillDown(HRotate(DRotate($T''^{VENTES}$, *TEMPS*, *AGENCES*, *ens*), *VOYAGES*, *cla_int, cla_fr*), *VOYAGES*, *TypeV*)

**4. Conclusion**

Cet article définit un modèle de représentation des données en constellation pour les bases multidimensionnelles. Ce modèle assure une grande cohérence des données par sa propriété de multi-instanciation (conditions d'appartenance des instances des dimensions aux hiérarchies) et un ensemble de contraintes exprimant l'inclusion, l'exclusion, la simultanéité, la partition et la totalité entre les hiérarchies. Nous distinguons les contraintes intra-dimension qui portent sur les hiérarchies d'une même dimension, des contraintes inter-dimensions qui s'appliquent entre les hiérarchies de dimensions distinctes.

Nous avons étudié également l'impact de ces contraintes sur les principales opérations multidimensionnelles (cet aspect est peu étudié dans les travaux existants). Nous avons étendu les opérateurs multidimensionnels afin de leur conférer une propriété supplémentaire offrant la possibilité de maintenir ou d'étendre les analyses en fonction des contraintes existantes entre les hiérarchies. Ces contraintes contribuent à améliorer la cohérence des analyses et facilitent la tâche d'analyse en précisant le domaine d'étude.

Nous poursuivons ces travaux par l'intégration des contraintes dans le langage assertionnel OLAP-SQL (Ravat *et al.*, 2002) que nous développons. Ce langage permet actuellement de définir, de manipuler, de consulter et d'interroger les données entreposées dans une base en constellation. L'ajout de contraintes offrira un ensemble de fonctionnalités avancées pour la configuration des bases en constellation ; le langage de définition devra notamment être complété par un langage de définition des contraintes.

## 5. Références bibliographiques